# A Synchrotron as Accelerator of Science Development in Central America and the Caribbean

GALILEO VIOLINI, VÍCTOR M. CASTAÑO, JUAN ALFONSO FUENTES SORIA, PLÁCIDO GÓMEZ RAMÍREZ, GREGORIO MEDRANO ASENSIO, EDUARDO POSADA, CARLOS RUDAMAS

ABSTRACT

**Purpose**: We want to identify possible ambitious projects that, through a strong regional collaboration, may accelerate the process of filling the scientific gap between Central America and the Caribbean (CAC) and the rest of Latin America.

**Design/methodology/approach**: We analyze the actual mechanisms used in the past for the above purpose, in order to identify the bottlenecks at the origin of the problem, and analyze possible experiences in other regions that have fulfilled analogous results.

**Findings**: The main bottlenecks are the scarcity of public research centers and little or no research in private universities. This requires that the "Central American Science and Technology Fund" proposed by the government of Guatemala, and pending of a decision by the Summit of Central American Heads of State be approved, and focus on objectives capable of attracting the attention of the non-academic sector, first and foremost policy makers, but also civil society in general. This requires ambitious projects, and the successful experience of SESAME ("Synchrotron Light for Experimental Science and Applications in Middle East ") offers an interesting basis for reflection, as it allows scientific research and short-term practical and social applications. Such an objective may attract the attention of the non-academic sector, first and foremost policy makers, but also civil society in general. The understanding of the value of the initiative by these sectors is crucial to create the conditions for the necessary financing by the governments of the region, possibly in collaboration with the regional and international development banks. The foreseen outputs will be both of academic and social impact. Beside the direct use of the synchrotron, one can expect advantages for doctoral training and the prevention of brain drain. More generally, as David Gross uses to remind: Science drives Technology, Technology drives Innovation, and this eventually leads to the upgrade of the welfare.

**Originality**: A regional synchrotron may be the first historical example of a big international facility in the Great Caribbean Region, that does not have any of this level. Moreover, both for the training of the to-be users and for the implementation of the project, South-South cooperation may be useful and this may favor the interest of international donors to consider new mechanism and goals for their international cooperation with the region, and strengthen it, after the stalling of the last decades.

## 1. Introduction

In the last 40 years the economic importance of Latin America has decreased. Its contribution to World GDP has lost two percentual points, but this does not tell the full story. The per capita GDP, that used to be significantly higher (about 12%) than the World average, is now significantly lower, by the same percentual difference. On the basis of these World Bank data, it has been suggested that this is the result of structural socio-institutional environment and of the neoliberal economic policy of the region, that are cause and effect of what is called the vicious circle of underdevelopment. The conclusion is that "today Latin American countries face the challenge of radically reorienting their development policy and strategy" and "they should use as a point of reference the successful experience of the emerging countries from Asia". (Sánchez-Masi. 2021)

The most famous experience is the South-Korean one. This country, partly because of favorable geopolitical conditions, but also thanks to a very successful development policy, succeeded in increasing its GDP per capita (less than 1 dollar at the independence, and still 100 US $, in 1960), exponentially up to more than 11000 US $, level reached just before the Asian financial crisis of 1997, and since then, despite a few minor occasional negative effects, is now more than 31000 $. This was the result of a state policy in which Science played a major role (Seung-Hun Chun, 2010).

Indeed, it is broadly recognized that Science plays a key role in the development of nations. In particular, Physics, that has been instrumental to the global technology advancement in the last 100 years in the World, is less developed in Central America than in the rest of Latin America. (Dutrénit *etal.*, 2021), This is confirmed by periodical analyses (Wray, 1967; Mora, 1993; Violini, 1993; García Canal, 2011) that, despite exhibiting an undeniable progress, call for an acceleration in the scientific policy of the region, noting the significant gap with respect to the rest of Latin America.

A reorientation of national and regional science and technology policies in favor of basic science in general may certainly be necessary and useful, but, without the courageous decision of committing the region in ambitious projects, future analyses will continue to register the same situation. Zeno's slow progress is not the solution, rather a Copernican revolution is needed.

Accordingly, the purpose of this note is twofold: to show that this revolution is possible and only requires a political decision and to provide to policy-makers and civil society the background information they may need to make or support such decision.

## 2. Science in Latin America and Great Caribbean

The analysis and proposal of our note focuses on the Central American countries and those others that constitute the Caribbean region, especially the biggest islands: Cuba, Dominican Republic, Jamaica and the Free-Associated State of Puerto Rico, with a complex political connection to the United States. In the following we shall refer to this broad region as Central American - Caribbean (CAC).

The CAC region, which has a population of around 60 million, is rather homogeneous in terms of social, economic, scientific and technological development. Central America in particular has a population of about 50 million that are divided between seven independent nations: Panama, Costa Rica, Nicaragua, Honduras, El Salvador and Guatemala (which were all previous Spanish colonies) and Belize, a parliamentary constitutional monarchy associated to the United Kingdom. Geographically speaking, as we shall do in this note, one refers to the Great Caribbean as the relatively extensive region that includes, beside both Hispanic and Anglo-Saxon independent countries, as well as portions of South American countries, such as Colombia, and North American, such as Mexico.

There are only small differences in terms of economy and social development between Central American and Caribbean countries, which is evidenced by the fact that almost all are classified in the group of developing or middle-income countries, and none belong to the select group of nations that fall under a high-ranking Human Development Index). It is even noted that differences of an order of magnitude in GDP per capita are not reflected in large differences in the HDI, which in SICA countries varies between 0.815 in Panama and 0.634 in Honduras (Expansion, 2021a). A separate case in the region would be Haiti, whose GDP is 30% lower than the country with the lowest GDP per capita, Nicaragua, and an HDI of 0.510. (Expansion, 2021b)

One can notice that middle-income developing countries are growing but often not highly resourced or productive in relative terms. International attention tends to focus either on the group of more population and productive emerging economies, or on the group of extremely poor countries that need to receive international assistance, such as, in the region, Haiti. Instead, this group neither benefits from the trade gains of emerging economies, nor qualifies for significant donor aid. Therefore, for these countries it is critical to cultivate scientific and technological resources to advance national development goals. (Svenson, 2013).

After the second half of the past century, all CAC countries have achieved some level of scientific development thanks to the implementation of specific policies. Although the main object of our consideration is Central America and the Caribbean, analogous situations exist elsewhere, so that this study can be useful also in such cases.

An important factor of this development that deserves to be highlighted is the institutional progress based on the establishment of entities that work towards scientific and technological development, such as ministries and programs that focus on the education of researchers, research infrastructure and financial funds for I + D + i projects.

These actions have been implemented taking advantage of a policy on regional integration that dates back to the late 40s, and has allowed the establishment of regional and inter-regional programs with diverse goals, for example, courses and doctorate programs and the creation of entities such as the Consejo Superior Universitario Centro Americano (CSUCA) and the Comisión de Ciencia y Tecnología de Centroamérica, Panamá and Dominican Republic (CTCAP). This initiative also went to the extent of the foundation of local funds that provided financing opportunities for scientific and technological research projects, as well as collaborative projects.

Some important results, related to engineering fields, can be mentioned, such as the G-CREAS or Regional Accreditation System for Engineering of the Greater Caribbean, which seeks validation of its processes by the Accreditation Board for Engineering and Technology (ABET) of the US, being both programs that focuse on the quality of technological academic programs (GCREAS, 2021; ABET 2021).

However, for what concerns basic science, this policy will produce effects only over a long term, with an eventual foreseeable result that a deep gap remains between CAC and advanced scientific countries, unless an acceleration of the process is put in place.

The acceleration of science is a matter not only of concern to CAC but also to developed countries. Examples of strategic alliances on scientific research and technological innovation, are the Common Research Area between the European Commission and CELAC countries, and the Partnerships for Enhanced Engagement in Research (PEER) program, administered by the NSF of the USA.

In Central America, between 1966 and few years ago, one of the few important regional activities in Physics has been the Central American Course of Physics (CURCAF: Curso Centroamericano de Física), promoted by Robert Narvaez Little (University of Texas at Austin), under the umbrella of the Central American Physical Society (SOCEF: Sociedad Centroamericana de Física). CURCAF's model was the Latin American School of Physics (ELAF: Escuela Latinoamericana de Física), created in 1959 by the mythical fathers of Latin American Physics, Juan José Giambiagi, José Leite Lopes and Marcos Moshinsky. CURCAF's main focus was the improvement of the teaching method of Physics at school level, much under the influence of the PSSC project, whereas ELAFs refocused attention on advanced research.

This reflected the consideration that the prerequisite of scientific literacy in a country is a good system of secondary education. For a discussion of the linkage between the two and of the inequalities that affect the region, without entering in details about this problem, we refer to (Salas, 2007).

The major organizations that contributed to the flourishing of both initiatives were the Organization of American States (OAS), UNESCO, and the intergovernmental Latin American Center of Physics (CLAF: Centro Latinoamericano de Física), created by UNESCO in 1962, that has recently established a regional Central American headquarters in Costa Rica (CLAF, 2021).

In the early 1980s, Dominican Republic joined CURCAF, and a second C, for "Caribbean" was added. This was one of the first participations of the Dominican Republic in the not always appreciated (Cifuentes, 2012) complex and variable-geometry process of regional integration (Brotóns, 2013). The regional integration of Dominican Republic would be strengthened in 2012, with its adhesion to the Higher Central American University Council (CSUCA: Consejo Superior Universitario Centroamericano) and in 2013, with that to the Central American Integration System (SICA: Sistema de Integración Centroamericana).

In the 1990s, in the framework of a program for the post-civil-war reconstruction of Universidad de El Salvador (UES), a conference to draw a regional plan for Science and Technology was held at UES (Vary & Violini, 1995). Instrumental were the visionary support of the UES Rector, Fabio Castillo, and the understanding of the program supervisors in Brussels.

At the turn of the new millennium, the need of launching Physics Masters programs in the region was addressed (García Canal, 2011).

However, the crucial issue is the training of PhD in the region, as had been stressed by Giambiagi in a rare paper of about thirty years ago, that indicates for the end of last century quantitative and qualitative goals for Latin American Physics still now not yet reached (Giambiagi, 1993).

Years later, CSUCA's secretary general, Juan Alfonso Fuentes Soria, and the fourth director of the "Abdus Salam" International Center of Theoretical Physics, ICTP, Fernando Quevedo agreed that two regional PhD programs, in Physics and Mathematics, were needed, and their creation was proposed (Quevedo *et al.*, 2012). The programs were prepared by mathematicians and physicists (holding a PhD degree) of universities members of CSUCA, along with advisors from North America, South America and Europe. This took place in several meetings organized by CSUCA in different locations in Central America. The doctorate programs were approved by CSUCA in 2012 (CSUCA, 2012), and started operating in 2014, with the Physics program at UES, the Universidad Tecnológica de Panamá (UTP) and the Universidad de Panamá (UP), and the Mathematics program at the UP. Despite that their start has not been without operational problems, they are most certainly a milestone in the challenging higher-education policy of the region (Rodríguez Peña, 2020), because of the harmonization of the programs and the recognition of credits and titles by participant universities.

## 3. Recent Developments of Science in CAC

The seed has been fruitful. Currently, the development of a network called Researchers Network for the Development of Natural Sciences (RCN: Red de Investigadores para el Desarrollo de la Ciencias Naturales) is being supported by the General Secretariat of SICA, CSUCA, the Commission for the Scientific and Technological Development of Central America, Panama and the Dominican Republic (CTCAP), and the International Development Research Center (IDRC), within the framework of the project "Strengthening science and innovation policy capacities in the member countries of the Central American Integration System". RCN is a network of researchers from Central America and the Dominican Republic, with international links, whose objective is to promote and strengthen regional doctorates in natural sciences and to promote the development of multi- and interdisciplinary research projects in natural sciences and related areas. This leads to a future of positive scenarios. RCN expected results are a strengthening of the Physics regional doctorate (Rudamas *et al.*, 2021), an extension to Biology and Chemistry (Nitsch *et al.*, 2021) and the fostering of a broad network of regional collaboration, with much less dependence on the goodwill of individuals.

There are also few national science-related doctoral programs. For example, the Doctorate in natural sciences for development (DOCINADE: Doctorado en Ciencias Naturales para el Desarrollo) in Costa Rica. The program is developed on an inter-university basis between the Tecnológico de Costa Rica (TEC), the Universidad Nacional (UNA) and the Universidad Estatal a Distancia (UNED). The PhD program offers the following emphasis and disciplinary areas: Agricultural Production Systems, Natural Resources Management, Environmental Management and Culture, and Applied Electronic Technologies. In the Dominican Republic there are PhD programs in Environmental Sciences and in Energy Management for Sustainable Development at the Instituto Tecnológico de Santo Domingo (INTEC) and also a PhD in mathematics, run by a public-private consortium of three Universities (Gómez, 2020), although it is unclear why no synergy with the regional CSUCA program has been sought.

These programs can reduce, although only marginally, the need for PhD studies abroad, that often end in brain drain (Violini, 1991), a problem that tends to be general, not only for the developing countries.

In the region, the mitigation of the effect of brain drain has been pursued by yearly scientific meetings, such as Converciencia in Guatemala (Chang, 2020) and the International Congresses of Scientific Research of the Dominican Ministry of Higher Education, Science and Technology (Gómez *et al.*, 2020). A similar program, the Encuentros Científicos Internacionales (ECIPERU, 2021) takes place in Peru. Other initiatives aim to take advantage of the scientific diaspora, such as the Argentinian RAICES network (RAICES, 2021), the ephemeral Red Caldas in Colombia, now limited to country clusters (Chaparro *et al.*, 2004) and the United Nations Volunteers (UNV)-Transfer of Knowledge Through Expatriate Nationals (TOKTEN) Program promoted by the United Nations Development Programme (UNDP).

## 4. Local Impact of Science

Despite these initiatives, local high-level training and utilization of the diaspora's expertise are just a first step. Their local impact is null, unless these trained scientists may find an adequate research environment and real opportunities. The bottleneck is that public research centers are few and private research almost inexistent. Thus, universities, often with heavy teaching schedules and small amounts of research, remain the almost unique available providers of employment, just as it was more than fifty years ago (Wray, 1967).

Two other mechanisms to foster regional scientific progress have actually been proposed. The first is the creation of regional Centers of Science (Violini, 2017), using the model of ICTP. Forty years back, Abdus Salam, who had just received the Nobel Prize for Physics, promoted the creation or development of "triestinos", i. e. Trieste-like, centers worldwide. It was a big international project that culminated in a Conference at ICTP Headquarters in Trieste and in Salam's unsuccessful attempt to receive financing for the initiative.

In the US, a triestino center was created by initiative of Marshall Luban and James Vary at Iowa State University. For almost a decade it played a complementary role to ICTP, especially in Africa. Moreover, at Fermilab, Leon Lederman and Roy Rubinstein, promoted collaborative programs with Latin America.

In Latin America, Salam's program included the Instituto de Estudios Avanzados, IDEA, in Venezuela, Multiciencias, in Peru (Latorre, 2018), the Centro Internacional de Física, CIF, in Colombia (Fog, 2011), and the Centro Internacional de Física y Matemáticas Aplicadas, CIFMA, in Mexico, this last one through Salam´s Deputy, Luciano Bertocchi (Flores Valdez, 2018). The last three centers had a close and strong interaction. CIF had a representation office in Lima (and another in Quito) and its experience was frequently shared with Mexico (Moore, 1983). Unfortunately, Multiciencias did not survive the Peruvian economic crisis of the so-called lost decades (Llosa & Panizza, 2015). Instead, CIF and CIFMA are still active, and CIF has been recognized by TWAS as a developing-country Center of Excellence (Fog, 2013).

In the last decade, the idea was brought back to life, thanks to Fernando Quevedo mentioned previously. With ICTP support, a few aligned Centers were created worldwide and recognized by UNESCO as Category 2 Centers. Two of them are in Latin America, the Mesoamerican Center for Theoretical Physics, MCTP, in Mexico, and the South American Institute of Fundamental Research, SAIFR, in Brazil. The former includes in its mission collaboration with CAC and the Andean region. In the Andean region, similar attempts took place in Colombia and Ecuador, and in the Caribbean, where a center on Materials Science, CRICMA, has been proposed (Núñez Sellés *et al.*, 2018). Despite initial enthusiasm, their implementation failed for different reasons that could be interesting to analyze, even if it goes beyond the purpose of this note.

Obviously, Physics research by individual groups is also carried out, and regional networking is growing, thanks to the existence of very high-level niches in most CAC countries, and to the RCN experience (Rudamas *et al.*, 2021). This is not limited to groups with a strong historical regional vocation, but also reflects policies of excellent institutions like the Pontificia Universidad Católica Madre y Maestra (PUCMM), with its projects of a Master Program in Physics and a doctorate in Materials Sciences.

However, in countries that fund science with less than 1 % of their GDP, (in some cases even less than 0.1 %), the first most pressing need is to effect significant increases in this financing. This may happen through an instrument originally proposed in a CSUCA doctorates meeting, that acquired consent when Guatemala's vice-president, Juan Alfonso Fuentes Soria, proposed it, on behalf of his government, to the 2015 Central American Summit. It consists on a Central American Fund for Science and Technology (Fuentes Soria, 2017). The proposal has been observed with interest, but the final decision for its creation, which is in the hands of the Summit of the Central American Heads of State, is still pending.

The purpose of the Fund for Science and Technology is not to increase the current national support to science and technology by some scaling factor, but rather to foster broad-scope regional initiatives, like doctorates or the establishment and management of advanced research centers, possibly stimulating a boost of the private sector participation.

## 5. Potential Venues

The central question, however, is this: would strengthening the doctorates, having research centers and having the Fund in place be sufficient to develop CAC Physics and Science at a pace comparable with that of other Latin American countries?

The plain answer is: most probably not!

The absence of interest and support, by the civil society, the press, the private sector and the political sector is notorious. Some of these problems are broadly recognized, although the limited development of Central America is also linked to the linguistic barrier caused from having Spanish as broadly common language (Svenson, 2013). Incidentally, our proposal goes in the direction of overcoming this problem.

In a webinar organized by the Konrad Adenauer Stiftung in Guatemala, on the Challenges for the region (Konrad Adenauer Stiftung, 2021), the emphasis was on economics, trade, politics, problems like health, education and food security (Padilla Vassaux, 2021), but missing the necessary condition for any sensitive policy in these areas, a solid scientific culture of the region.

Certainly, some interesting ideas still appear. For example, the Dominican president, Luis Abinader, announced the intention of creating a Silicon Beach (Abinader, 2021), a city of knowledge, to some extent similar to that put in place by Panama in the former Canal Zone, that has contributed to a remarkable scientific development in the last two decades. There is no doubt that such a proposal may have a tremendous impact for the development of the country and of the entire region (Violini, 2021). However, six months after that announcement, no concrete actions have been made, nor was it mentioned again at president Abinader's presentation of the first-year results and future programs of his government (Abinader, 2021).

It must be stressed that the purpose of establishing a regional Fund is not to allow politicians to claim that their plans of country development do not neglect Science. The mechanism will not produce significant effects, unless it is used to allow the region to realize the ambitious programs it needs that cannot be realized under the current financing conditions.

The success of the Fund will be measured by its capacity to implement such programs and in a reasonably short time frame. It is a goal that cannot be limited to Central America and Spanish-speaking countries. Caribbean members of Caricom and the Association of Caribbean States, ACS, must also be involved.

Thus, the Fund will make possible a regional science policy focused on goals that may attract the positive attention of the non-academic sectors. Their support to the recognition that scientific and technological development must be a state priority is a necessary condition to make possible such a change of paradigm. If this happens, it is foreseeable that the regional development banks operating in the region, and whose crucial importance for the regional development has been stressed in the mentioned webinar (Deras, 2021), may play a more active role, in synergy with the broader-scope Inter-American Development Bank and World Bank.

However, what could be such an ambitious program? The relation between science and society is complex, and, unfortunately, it has been usually analyzed through the lenses of advanced countries.

A feature of R&D in developing countries, which is especially evident in Central America is the prevalence of applied research, reflecting the prejudicial idea that they must limit themselves to socially and/or economically useful science, idea that is not so appealing anymore. The well-known sentence attributed to Faraday about taxes that would come from apparently non-applicable research applies to them, as well. Another difference with respect to advanced countries is in the larger weight of state financing of the research (Giambiagi, 1993).

Approximately 15 years ago, Jean-Paul Fitoussi warned that "a naive conception of research aimed to obtain quick results may prevent financing fruitful projects if they do not appear susceptible of practical applications. … This may be normal for socially efficient public policies, but it risks to ignore important projects" (Fitoussi, 2005). About 15 years before that, Galileo Violini and Marcello Cini, the coauthor of "The bee and the architect" (Ciccotti *et al.*, 1976), bible of the neo-Marxist approach to the relation between science and society, had reflected on certain peculiarities of this relation in developing countries (Cini & Violini, 1983), and this favored, to some extent, the Andean participation in Fermilab, and paved the way to a collective study CIF promoted ten years later, in collaboration with CERN, Fermilab, ICTP and TRIUMF (Aguirre *et al.*, 1993). The collaboration with CERN developed ten years later, linked to the HELEN project, (Maiani, 2005) among whose promoters, beside Luciano Maiani and Verónica Riquer, was one of members of the aforementioned Steering Committee, Juan Antonio Rubio.

However, Cini and Violini referred to small investments intended to allow to satisfy personal research vocations. If ambitious and expensive national projects are at stake, one cannot ignore, as Fitoussi recognizes, their cost of opportunity, in particular, in countries with great problems of social equality, and this applies especially to the present times, because of the economic impact of COVID pandemic.

## 6. A specific Proposal

In this context, the successful experience of SESAME, (Synchrotron Light for Experimental Science and Application in the Middle East) offers interesting elements to reflect on. Beside its scientific value, it also satisfies the requirement of practical and socially useful applications, and this happens in countries of economic capacity and scientific tradition-culture not much different from CAC.

At the end of last Century, when Germany was going to dismantle its Bessy-1 synchrotron and to put in place a new one, SESAME started the proposal of relocating Bessy-1 in Middle East. The idea received a strong support from Herwig Schopper (CERN) and Herman Winick (Stanford). Jordan was chosen to be the host country for SESAME. The limited specific experience of the scientists in most participating countries was not a problem because the time required to start operations was used to realize a strong several-year training program. Eventually, SESAME opened in 2017 (Blaustein, 2017) and now it is a successful regional laboratory (Aleisa *et al.*, 2021).

In fact, three beamlines have been commissioned, three more are under construction, with plans for a seventh. Demand is high, with 151 proposals between the first call in December 2016 and the third in November 2019. Between July 2018and February 2020, experiments have been conducted for 62 of them, involving 12 different countries, many of them collaborative projects. Its outreach activity included hosting another organization's workshop on its premises, that of the Association of Arab Universities. SESAME's solar power plant, inaugurated in February 2019, makes SESAME the world's first large accelerator complex to be fully powered by renewable energy and the world's first carbon-neutral accelerator laboratory (Aleisa *et al.*, 2021).

A few years ago, there were about 60 synchrotrons in the world (Lightsources, 2015), but now there are a few more. About one third of them are located in the American continent. Among them, only two are South of Rio Bravo, the physical, non-anthropic, border between US and Mexico. They are both in Campinas, Brazil. One is the LNLS, built in 1997, and the other a fourth-generation synchrotron, SIRIUS, inaugurated in 2018. As a comparison, the energy of LNLS is 1.37 GeV, and that of SIRIUS, whose radius is about five times larger, 3 GeV.

This makes it appropriate to recommend that having at least one more synchrotron must be recognized as a priority in Latin America. One could wonder whether for this it may be convenient to relocate the old Brazilian machine. But this is a detail and cannot be **THE** reason whether to put in place or not a synchrotron as we propose.

Latin America deserves it anyway.

The variety and versatility of synchrotron capacities may help overcome the need of sending students and young scientists abroad. Synchrotrons can be used for biological, chemical and physical research. A short list of possible applications includes structural molecular biology, molecular environmental science, surface and interface science, micro-mechanical devices, x-ray imaging, archaeological microanalysis, materials characterization and biomedical applications.

We recall that, while the possibility of a large-scale production radio isotopes (RI) from photons was considered very unlikely in 2009 (IAEA, 2008), now -- at least at the proof-of-concept level-- it seems highly probable, because of the great progress in the use of electron accelerators and synchrotrons (Ejiri & Date, 2011). This suggests that this might be the dominant method for RI production in the next decade (Srivastava *et al.*, 2021).

Therefore, such a facility will not only benefit Physics. It will benefit Science in general, contributing to scientific development and to tackle and solve many national and regional problems, but there are also other aspects to underscore.

## 7. Concluding Remarks and Expectations

The availability of such a facility will have impact on the training of graduate students, postdocs, and other young scientists, who will no longer have to go abroad to pursue their scientific interests. This has been the case in countries such as Taiwan and South Korea, where national facilities built many years ago have significantly reduced brain drain, and encouraged their scientific diaspora to return home. Of course, a word of caution may be opportune. The building of a new facility, and this happened for example in Trieste, when ELETTRA, the synchrotron of the Trieste Research area, was built, or in Frascati in the 60's, triggers a surge in physics enrollments, which may last for quite a while. Obviously initially this is viewed as "good", at least by the physics department itself. After all, especially these days, a doubtful (Ordine, 2021), neoliberal justification of the role of the universities may be based on indicators of this sort. It is obvious that the facility can only absorb a fraction of the increased number of local graduates in physics, and this may eventually lead to a problem of internal brain drain, unless its creation is accompanied by a development of the demand of graduates. However, in the specific case of CAC, where a large number of universities do not offer physics careers, and not even have Physics departments, it is likely that this risk is reduced.

The construction of a synchrotron will stimulate local industries, including services, suppliers of all sorts, new high-tech companies attracted to the site, etc. Some of these would later take advantage of its capacity. Qualified technicians, highly needed in the region and in the rest of Latin America, will be trained, and this can be of the utmost usefulness in other branches of research, like Astrophysics, an area of growing importance in the continent, even if not particularly in CAC (Hiroaki *et al.*, 2021), as well as in several industrial activities that require such expertise, thus opening opportunities for job creation and industrial development.

Another important perspective is Scientific Diplomacy. SESAME, with a membership including Cyprus, Egypt, Iran, Israel, Jordan, Pakistan, the Palestinian Authority and Turkey, is a proof of the capacity of science to promote the dialogue and to realize the goal UNESCO states in the preamble to its constitution "to construct the defences of peace in the minds of men and women". Spanish-speaking Central America has been a leader in regional integration, but still much has to be done for a Great Caribbean integration. This project can contribute to this goal.

It is foreseeable that if the project of a CAC synchrotron materializes, it may count on a vigorous international cooperation. At a scientific level, this happened for SESAME, and we would dare to believe that SESAME will certainly be an enthusiastic supporter of an initiative born from its experience. ICTP can be instrumental to receive support from ELETTRA. We also have reasons to believe that a valuable support will come from the Brazilian and some US synchrotrons.

Another initiative that may be useful for this purpose is LAAAMP (Utilization of Light Source and Crystallographic Sciences to Facilitate the Enhancement of Knowledge and Improve the Economic and Social Conditions in Targeted Regions of the World) (LAAAMP, 2021), It was launched in 2017 as a cooperative effort between IUPAP and IUCr to promote synchrotron radiation and crystallographic sciences in Africa, Mexico, the Caribbean, Southeast Asia, Middle East and the Pacific Islands. The project was supported by a grant of the International Science Council (ISC). LAAAMP focuses on training young researchers, reaching out to professionals such as university faculty, and engaging the public and governmental officials in discussions about the role that advanced light sources and crystallographic sciences could play to improve their countries' educational institutions, economies, social structures, health and world competitiveness.

It is necessary that these possible sources of support materialize as soon as the project is approved, in order to start a strong program of training. In fact, the need to ensure the operation and maintenance of the facility requires the launch of a large-scale program of human resources training, as it happened for SESAME. This training may require a few-year program, and possibly some conversion of theorists to experimental physics. This would be a courageous, not always easy, personal choice, but this is how High-Energy experimental Physics developed in Mexico, with Clicerio Avilez, and in Brazil, when Jayme Tiomno, in a conversation at Fermilab´s cafeteria witnessed by one of the authors (GV), convinced Carlos Escobar and Alberto Santoro to make this step. Specifically in the area of electron accelerators, one can find an example of great results and motivation to follow, that of Bruno Touschek, a theoretical physicist, father of the intersected rings with Frascati's AdA (Amaldi, 1981; Pancheri, in press; Bonolis, & Pancheri, 2011).

Among some valuable experiences in developing countries, we highlight the cases of the African Light Source, AFLS (AFLS, 2021), which includes conferences, schools, training and mobility through a Consortium in that continent, and of an analogous project in South-East Europe (Šlaus, 2021). Not only they stress further the need to keep the pace in CAC, but also offer a perspective for a coordinated effort to carry on jointly the required training activity with the possibility of finding external support in suitable European, American or Chinese programs. For this an important role can be played by LAAAMP and by cooperation organizations like DAAD (Deutscher Akademischer Austauschdienst) that traditionally have supported South - South cooperation.

Cooperation is crucial in terms of the cost of this project. This would depend on many variables, and in this note we avoided a deep discussion of this point because it could be a distraction from the central issue. Nevertheless, this cannot be a problem once the political will exists. Because of the COVID-19 pandemic, the Central American Bank for Economic Development gave to each SICA country 50 M US$, whose total is more than what would be needed for any reasonable choice of synchrotron characteristics.

Foreign support may come from cooperation-for-development funds. Over the last 20 years, their amount for Central America has been stable around 1 billion dollars, (which in real terms means a decrease of about 35%). Current Direct Foreign Investment in the region is of the order of 15 billion dollars (Deras, 2021). The broadening of the scope of the Association Agreement between Central America and the European Union, from its original vision limited to its economic and commercial impact (Zabalo *et al.*, 2019), and the European Union program of support for CELAC infrastructure could be suitable instruments to provide this project international matching funds to an investment coming from the region. The relevance of the project certainly makes it eligible for an IDB support through its Regional Public Goods program. Finally, if, following Germany's step, it is decided to reinstall the dismantled LNLS in another country, it could be a possibility for Brazil to bear the corresponding costs, either at federal or at São Paulo state level.

These considerations confirm that the feasibility of this project does not critically depend on money.

It may be good to recall a lost-opportunity experience of the region. Forty years ago, the first CIF (actually ACIF) activity was a high-level workshop on Gravitational waves (Posada & Violini, 1983). Unfortunately, it did not lead to the expected result of catalyzing the birth of a research group that could participate in the detection, on February 11, 2016, of gravitational waves (Abbott *et al.*, 2016). This detection was the result of an international effort of 94 research groups, among which Latin American institutional representation was limited to two groups from Brazilian institutions, one of which is the afore-mentioned SAIFR, although one of the spokespersons who made the announcement was Gabriela González, an Argentinian scientist working in a US university (LSU, 2016). The majority of the Bogota speakers could not be among the authors of (Abbott *et al.*, 2016), but one of them, Fulvio Ricci, did, as one of the leading Italian participants in the discovery.

What could have been the impact on the Science in all of Latin American region, if an Andean group generated by the visionary CIF workshop had contributed to that historical result?

Last June, opening the XVI Dominican Congress of Scientific Research, David Gross advocated for the importance of research excellence, recalling that Science drives Technology, Technology drives Innovation, and this has a final result in society´s welfare. (Gross, 2021)

A regional Synchrotron can be the way to make this real in the CAC Region, with the historical significance of being the first example of a big regional facility there.

How can we ensure that these considerations receive the due attention by the civil society, the mass media, and the private sector?

The future will tell whether the scientists of the region will succeed to attract that attention obtaining a thorough diffusion of the project and that the debate is not limited to the academic milieu, but we would like to recall the prescription of a leading Tanzanian scientist and member of the UN-SG-Scientific Advisory Board, Abdallah Daar, during the 2015 TWAS Meeting in Oman: “Make a long-term commitment. Connect research to social value. And don’t be afraid to fail” (Daar, 2015).

If this happens, hopefully it will lead to a political decision that will determine whether, in the next decades, CAC countries will have the place they deserve among those that “participate in the progress and the movement of natural sciences” (Sarmiento, 1871).

## FINAL REMARK

After this proposal was uploaded in arXiv (Violini *et al.*, 2021), we became aware of an analogous proposal of some years ago in Colombia (Gómez Moreno, 2014). That article provides a nice and accurate presentation of the theory, functioning, technical problems and needed available experience to install a synchrotron. There is no need to say that, if the conditions to implement such a project existed then, a fortiori they exist now.
When one looks at the scientific motivations of the proposal and at the non-scientific benefits Gómez Moreno foresees, that to a very large extent coincide with those indicated in this paper, it is impossible not to think that the region has lost once more an opportunity of progress and economic and social development. Now, this opportunity appears again. We hope that it will receive the consideration that unfortunately Gómez ‘s proposal did not receive.

### Acknowledgments

We had many interesting, stimulating discussions, and/or encouraging comments, and/or critical reading of the manuscript, with several colleagues and friends. We would like to especially thank for their interest and contribution: Jorge Cuadra (UES), Denia Cid Pérez, (PUCMM), Catalina Curceanu, (INFN), Mustafa El Tayeb, (UNESCO’s Science Sector former director), Ernesto Fernández Polcuch, (UNESCO’s Lima Office director), Emilia Giorgetti, (CNR), Adam Kaminski, (Iowa State University and Ames Lab.), Reza Mansouri, (former president, Iranian Physical Society), Darwin Muñoz, (Dominican national contact point of the European STI Commission), Joe Niemela, (ICTP and APS), Federico Rosei, (INRS), Rinaldo Santonico, (emeritus professor, University of Tor Vergata), Sandro Scandolo (IUPAP and ICTP), Ivo Šlaus, (honorary president, the World Academy of Art and Science), Yogendra Srivastava, (emeritus professor, Northeastern University), Harry Westfahl Jr., (LNLS director), Herman Winick, (SLAC Emeritus Professor), and Moneef Zou’bi, (co-founding director, World Sustainability Forum).